\documentstyle[prl,twocolumn,aps,psfig]{revtex}

\def\putfigl#1#2#3{\setbox20=\hbox
	{
	\psfig{file=#1,height=#2cm,angle=#3}
	}
	\centerline{$\vcenter{\box20}$}}
\begin{document}
\tighten
\title{Bending  and Base-Stacking Interactions 
in Double-Stranded Semiflexible Bioplymer}
\author{Zhou Haijun$^{1}$ and Ou-Yang Zhong-can$^{1,2}$}
\address{$^1$Institute of Theoretical Physics,
 Academia Sinica, 
P.O. Box 2735, Beijing 100080, China\\
$^2$Center for Advanced Study, 
Tsinghua University, Beijing 100084, China}
\date{Resubmitted on Nov. 10, 1998}
\maketitle
\begin{abstract}
{\bf Simple expressions for the bending and the
base-stacking energy of double-stranded semiflexible
biopolymers (such as DNA and actin) are derived.
The distribution of the folding angle between the
 two strands is obtained by solving a Schr\"{o}dinger
equation variationally. Theoretical results
based on this model on the extension versus
force and extension versus degree of supercoiling
relations of DNA chain are in good
agreement with the experimental observations of 
Cluzel {\it et al.} [Science {\bf 271}, 792 (1996)],
Smith {\it et al.} [{\it ibid.} {\bf 271}, 795 (1996)],
 and
Strick {\it et al.} [{\it ibid.} {\bf 271}, 1835 (1996)].}
\end{abstract}

Recent experimental developments  make it possible to
investigate directly  the mechanical properties 
of  single macromolecules, and it is shown
 that many biopolymers should
be regarded as semiflexible rather 
than as highly flexible \cite{r1,r2}.
The simplest and yet the  most famous model for a semiflexible  polymer is the
wormlike chain (WLC) model, it  can well describe
  the elasticity of DNA at the low
and the moderate  force ranges \cite{r2}.
 The WLC model 
regards a semiflexible polymer as  an 
{\it inextensible} mathematical
line characterized by a single length scale, the bending persistence length
 $\ell_p$ \cite{r2,r13}.
 However, many biologically important
 macromolecules,
including DNA and proteins such as actin,
are double-stranded.  
They are formed
by two  individual  chains  ($``$back-bones") bound 
toghther by permanent bonds ($``$base-pairs") such
as the hydrogen bonds in the case of DNA
 to achieve a ladderlike architecture.
Very recently, experiments on DNA and actin
 indicate that these
interwound structures  lead to  non-WLC elastic
 behaviors \cite{r3,r4,r5,r6,r7}.
To interpret these observations, 
 Everaers {\it et al.} \cite{r8}
 have made the first attempt to
regard a double-stranded (DS-) polymer
 as a polymer with
 two individual
wormlike chains. Later Liverpool {\it et al.}
 \cite{r9} extended this work  to
the three-dimensional case and
 investigated the statistical mechanics 
 of DS-polymers by mean-field treatments.
Nevertheless, at the present time, 
 a  mathematically
rigorous model for DS-polymers is still lacking and 
the rich phenomena observed in experiments
\cite{r3,r4,r5,r6,r7} are not
convincingly explained
\cite{r10,r19}. This situation is unsatisfactory.

Based on  the early efforts of Refs. 
\cite{r8} and  \cite{r9}, in 
this Letter we have developed a model for semiflexible
DS-polymers.
A simple bending energy expression is derived and the
folding angle distribution is obtained by solving a
Schr\"{o}dinger equation variationally. The present
model, after taking into account base-stacking
 interactions,
is applied to DNA and the derived theoretical results 
are in considerable agreement with
three  different groups of experimental observations
\cite{r3,r4,r5,r6}.

\begin{figure}
\putfigl{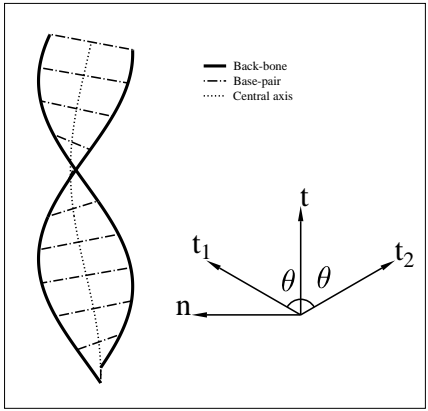}{7.5}{0}
\smallskip
\caption{(Left) sketch of a double-stranded polymer.
(Right) the definition of $\theta$ on 
 the local ${\bf t}$-${\bf n}$ plane, 
(${\bf b}={\bf t}\times {\bf n}$ is perpendicular
to the ${\bf t}$-${\bf n}$ plane). }
\label{fig1}
\end{figure}

 Mathematically, a  DS-polymer  is composed of 
two inextensible back-bones,
 $
{\bf r}_1(s)=\int^s{\bf t}_1(s^{\prime })
ds^{\prime }
$
and
$
{\bf r}_2(s)=\int^s{\bf t}_2
(s^{\prime })ds^{\prime },
$ 
both with the same
bending rigidity 
$
\kappa =k_BT\ell _p
$
\cite{r9}.
(Here the unit vector 
$
{\bf t}_i
(s)
$ 
$
(i=1,2)
$
is the   tangent of the back-bone 
$
{\bf r}_{i}
$
at its arc length $s$.)
Back-bones 
${\bf r}_1$ and  ${\bf r}_2$,
connected by many base-pairs of fixed length 
$2 R$,
are constrained to form the edges of a space 
ladder of fixed width, with 
${\bf r}_2(s)= {\bf r}_1(s)+2R{\bf b}(s)$
[Fig.\ \ref{fig1}]. Here 
${\bf b}(s)$
 is a unit vector on the ladder surface 
pointing from 
${\bf r}_1$ to ${\bf r}_2$ 
along
the direction of the base-pair,
 it is perpendicular to both edges
of the ladder, i.e., 
${\bf b}(s)\cdot 
{\bf t}_1(s)={\bf b}(s)\cdot {\bf t}%
_2(s)\equiv 0$ \cite{r9}.
 The central axis of such a ladderlike
 structure is defined
by ${\bf r}(s)={\bf r}_1(s)+R {\bf b}(s)$. 
We define the tangential (unit) vector of
the central axis to be ${\bf t}(s)$
 and ${\bf n}(s)={\bf b}%
(s)\times {\bf t}(s)$ to be a unit vector 
perpendicular to the ladder
surface.  It is useful to define the rotation 
angle from ${\bf t}%
_2(s)$ to ${\bf t}_1(s)$  to be $2\theta$
 (${\bf b}$ 
being the rotational axis, see Fig.\ \ref{fig1}).
 $\theta$,
 we name it the {\it folding
angle}, is in the range between $-\pi/2$ 
and $\pi/2$. From Fig.\ \ref{fig1}
we know that ${\bf t}_1=\cos\theta{\bf t}+\sin\theta{\bf n}
$ and ${\bf t}_2=\cos\theta{\bf t}-\sin\theta{\bf n} $. It is not difficult
to verify that \cite{r11}
\begin{equation}
d{\bf b}/{d s}=[{\bf t}_2-{\bf t}_1]/2 R
=-[\sin\theta/R] 
{\bf n}.  \label{eq1}
\end{equation}
Equation (\ref{eq1}) reveals a very important relation
 between ${\bf n}$ and $%
{\bf b}$ and it enbodies the rigid constraints
 that the two back-bones are
inextensible and that base-pairs of fixed 
length are formed between them.
With Eq.\ (\ref{eq1}) 
and the definition of ${\bf r}$ it 
can be  shown that \cite{r11}
\begin{equation}
d{\bf r}/d s=[{\bf t}_1+{\bf t}_2]/ 2= {\bf t}\cos\theta,
\label{eq2}
\end{equation}
i.e.,  $\cos\theta$ measures  the degree of
folding of the back-bones on the central axis. 

The deformation of the wormlike back-bones leads to the 
total bending energy of  the
system \cite{r9}, i.e., 
$$
\begin{array}{ll}
E_{\mbox{bend}}&=(\kappa/2)\int
 [(d{\bf t}_1/ds)^2 +(d{\bf t}_2/ds)^2 ] ds\\
\; &=\kappa\int [(d^2{\bf r}/d s^2)^2+
 R^2 (d^2{\bf b}/ds^2)^2]ds.
\end{array}
$$
Inserting Eqs.\ (\ref{eq1}) and (\ref{eq2}) into this 
expression we obtain an important result: 
\begin{equation}
E_{\mbox{bend}}=\kappa\int [ (d{\bf t}/ds)^2+
(d\theta/ds)^2 +\sin^4\theta/R^2] ds.  \label{eq3}
\end{equation}
We see that Eq.\ (\ref{eq3})
 is very simple as well as with
clear physical meaning: The bending
energy of the system can be seperated into two {\it independent} parts,
 the energy caused by bending
of the central axis and that
associated with the folding of the  back-bones.
  This observation
is very useful for our following discussions. 
(We notice  that the variables ${\bf t}$ and $\theta$
themselves  are 
 yet not completely 
independent of each other, because of Eq. (\ref{eq1})
\cite{r12}.) 

\begin{figure}
\putfigl{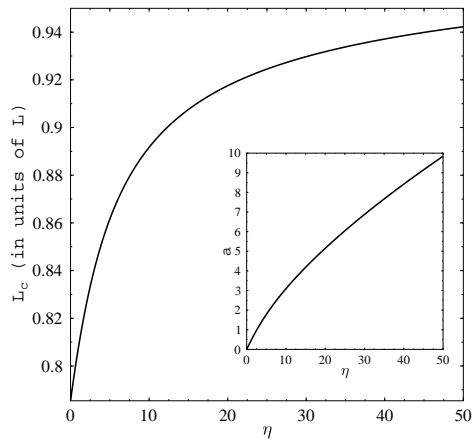}{7.5}{0}
\smallskip
\caption
{
The relation between  $L_c$ and $\eta$ according to
 Eq.\  (\ref{eq8}). (Insert:
 the variational relation between $a$ and $\eta$.)
}
\label{fig2} 
\end{figure}

A quantity of
immediate interest is the distribution function of
the folding angle $\theta$ \cite{r14}. 
Because of the energy expression Eq.\ (\ref{eq3}),
the $\theta$ distribution  $\Psi(\theta, s)$
 is governed by the following
Schr\"{o}dinger equation \cite{r2,r13}: 
\begin{equation}
{\frac{\partial \Psi }{\partial s}}
=-{\frac 1{4\ell _p}}\left( 
-{\frac{\partial ^2}{\partial \theta ^2}}
+4\eta \sin ^4\theta \right) \Psi ,
\label{eq6}
\end{equation}
where $\eta =(\ell_p/R)^2$ is a dimensionless constant.
Since $\theta \in (-\pi/2,\;\pi/2)$, the eigenvalues
of Eq.\ (\ref{eq6}) are all positive. Therefore,
the $\theta $ distribution 
is governed mainly 
by the ground eigenstate,  $\Psi_{gr}$, 
of Eq. (\ref{eq6}).
 However, the  eigenstates of 
Eq.\ (\ref{eq6}) cannot be
 obtained exactly and we have to
 take a varational approach by assuming $\Psi
_{gr}\propto \cos \theta \exp (-a\sin ^4\theta )$
\cite{r11}.
 Here $a$ is a
variational parameter to be determined  through
 minimization of the 
variational free energy density 
$
\int[ (d\Psi _{gr}/d\theta)^2+4\eta \sin ^4\theta
\Psi _{gr}^2 ] d\theta/(4\ell _p\int\Psi _{gr}^2 d\theta)
$ \cite{r2}.
The calculated relation between $a$ and $\eta$ is
shown in  Fig.\ \ref{fig2} (inset).
 With the $\theta$ distribution $\Psi_{gr}(\theta)$
 known,  
the average contour length $L_c$ of the central axis,
  according to Eq.\ (\ref{eq2}),
is related to $\Psi_{gr}(\theta)$
via 
$
L_c=\int_0^L\langle \cos \theta (s)\rangle ds=
L\int cos\theta \Psi_{gr}(\theta) d\theta
 /\int\Psi_{gr}(\theta)d\theta,
$
i.e.,
\begin{equation}
L_c=L\pi a^{1/4} D(2a)/[\Gamma(1/4)-\Gamma(1/4,a)],
\label{eq8}
\end{equation}
where $D(2a)=\;_2F_2(1/4,3/4;1,3/2;-2 a)$
is the generalized hypergeometric function and
$\Gamma(1/4,a)$ is the incomplete Gamma function,
 the constant
 $L$ is the total contour length of the back-bones.
The   $L_c$ vs. $\eta$ relation
is calculated numerically and
 shown in Fig.\ \ref{fig2}.
  From Fig.\ \ref{fig2} we
find that the
total contour length of the system increases nonlinearly with $\eta$. For $%
\eta\sim 0$, it approaches the lower limit of $L_c=(\pi/4)L$; for $\eta\rightarrow
\infty$, it approaches the upper limit of $L$. Notice that $\eta$ is
proportional to $T^{-2}$ \cite{r9},
 so the average contour length is sensitive to
temperature variations. This property of DS-polymers
may have some biological significance \cite{r15}. 

With the model for double-stranded polymers proposed
above, it is tempting for us to study the elastic properties of 
DNA macromolecules \cite{r10,r19}. The
two back-bones of DNA are bound together by base-pairs
to form a right-handed double-helix \cite{r16}.
In this case, the folding energy caused by the base-stacking
interactions between adjacent base-pairs should also be
 considered.  In this article,  we suppose that this
base-stacking folding potential together with the
back-bone folding potential [the term $\kappa \sin^{4}
\theta /R^2$ in Eq.\ (\ref{eq3})] as a 
whole can be modeled as a Lennard-Jones interaction
$
U=\sum_{i} U_{i,i+1}=\epsilon
[(\delta/r_{i,i+1})^{12}-2(\delta/r_{i,i+1})^{6}]
$
\cite{r16}, where  $r_{i,i+1}$ is the distance between
 adjacent base-pairs i and i$+$1.
For our purpose, we  convert this summation  into
an integral form as a contineous function of 
$\theta$ \cite{r11}.
Then
 the total energy of such a system under 
applied force $f{\bf \hat{z}}_0$ and torque $\gamma$ is
written as \cite{r11}
$$
E_{\mbox{total}}=\int\left\{
\kappa ({d{\bf t}\over ds})^2+\kappa({d\theta\over
ds})^2 \right.
$$
$$
+{\epsilon\over r_0}
 \left[({\cos\theta_0\over \cos\theta})^{12}-
2({\cos\theta_0\over \cos\theta})^6\right]
$$
\begin{equation}
\left.-f\cos\theta {\bf \hat{z}}_0\cdot
{\bf t}\right\} ds-\gamma \pounds,
\label{eq4}
\end{equation}
where $r_0$ is the back-bone arc length between 
 adjacent
base-pairs and $\theta_0$ is the equilibrium 
folding angle of the
back-bones when there is no force applied,
$t_x$ and $t_y$ are the  components of 
${\bf t}$ on the plane perpendicular to
 ${\bf \hat{z}}_0$, the direction of the external force.
$\pounds$ is the linking number \cite{r17}
 of the DS-polymer system, the
total topological turns the back-bones twist on
each other.  
$
\pounds=(1/4\pi)\int (t_x d t_y/ds-t_y d t_x/ds)ds
+(1/2\pi R)\int \sin\theta ds
$ for highly extended a DS-polymer
\cite{r11,r17}.
We define the dgree of supercoiling, $\sigma$, to be
$
\sigma=(\pounds-\pounds_0)/\pounds_0,
$
where 
$
\pounds_0=(1/2\pi R)\int \sin\theta_0 ds
$
is the linking number of  a relaxed DNA
\cite{r5,r10}.

\begin{figure}
\putfigl{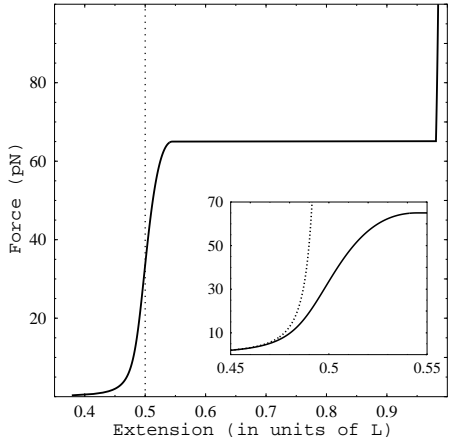}{7.5}{0}
\smallskip
\caption
{Force-extension curve for a
torsionally relaxed ($\gamma=0$) model DNA chain with
energy Eq.\ (\ref{eq4}).
 (Insert:  a part of this
curve, the dotted line characterizing
a  wormlike chain of length $L\cos\theta_0$ and
bending persistence length $2\ell_p \cos\theta_0$.)}
\label{fig4}
\end{figure}

First, we consider the case when $\gamma=0$, i.e.,
only external force $f$ but no torque is applied on DNA.
The theoretical results of the force-extension
relation for
such a model  DNA chain with
 energy expression Eq.\ (\ref{eq4})
are discussed below and   demonstrated in Fig. \ref{fig4}.

When $f$ is small,
 because of the strong folding interaction
the change  in $\theta$ is negligible
 and the DS-polymer is in a
double-helix configuration (whether
 right-handed or left-handed 
may be  determined by  other energy terms not
included here),
and  the polymer can
just be regarded as a wormlike
 chain of contour length $L\cos\theta_0$ and
bending persistence length $2\ell_p\cos\theta_0$
[Fig.\ \ref{fig4} (inset)].
 Therefore we focus our
attention to the  case when $f$ becomes high enough 
so the tangent
vectors ${\bf t}$ fluctuates around ${\bf \hat{z}}_0$ slightly.
In this case the equilibrium value of $\theta$ is 
related to $f$ via
\begin{equation}
f={12\epsilon\over r_0\cos\theta_0}\left[({\cos\theta_0\over\cos\theta})^{7}
-({\cos\theta_0\over\cos\theta})^{13}\right],
\label{eq10}
\end{equation}
i.e.,  the total contour length of the chain will
increase slightly with $f$.
The total extension is then calculated to be \cite{r11,r2}
\begin{equation}
z=\int^L_0 \langle {\bf \hat{z}}_0
 \cdot{\bf t} \cos\theta\rangle ds
\simeq\cos\theta\left(1-
{k_BT \over \sqrt{8\kappa f\cos\theta}}\right).
\label{eq11}
\end{equation}
However, the right-hand part in Eq.\ (\ref{eq10}) attains its
maximal value of
\begin{equation}
f_c\simeq  2.69\epsilon/(r_0\cos\theta_0)
\label{fc}
\end{equation}
at  
$
\theta_c\simeq \arccos(1.1087\cos\theta_0).
$ 
Therefore,
for $f >f_c$ the force equilibrium 
condition Eq.\ (\ref{eq10})
can no longer be satisfied
 and  the
back-bones will be fully extended, with $\theta\sim 0$. 
In other words, with the increase of external force,
a  double-stranded polymer will undergo an {\it abrupt}
 phase
transition from a double-helix (${\bf B}$-DNA) to a fully
 extended  ladder (${\bf S}$-DNA or
denatured DNA \cite{r3,r6,r10}) at force
$f_c$ [Fig. \ref{fig4}].
This phase transition for over-stretched DNA
 is indeed observed 
in lambda-phage DNA (see Fig.\ 2 in \cite{r3}
and Figs.\ 2, 4 and 5 in \cite{r4}, these figures are
very similar to Fig. \ref{fig4}).
To be quantitative, we set $\cos\theta_0=1/2$ (right-handed
helix), 
 $a=0.68\; \mbox{nm}$, $\ell_p=50\; \mbox{nm}$ 
\cite{r16,r1,r2}. Ref.  \cite{r4}
reports  $f_c$ to be about $65 \;\mbox{pN}$ at room
temperature, then we 
estimate that $\epsilon \simeq 600 k_B=2k_B T$ from our
theoretical formula Eq. (\ref{fc}). 
 This value of $\epsilon$ is
 in close agreement 
with the estimation of Strick {\it et al.} \cite{r6}.

\begin{figure}
\putfigl{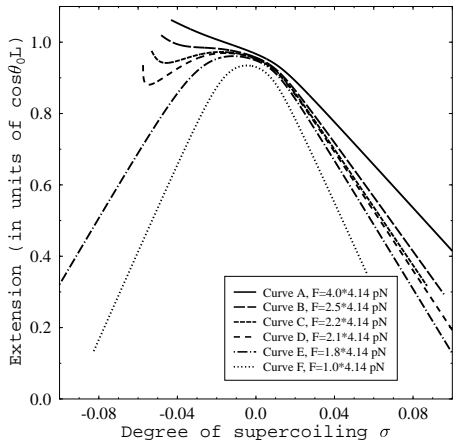}{7.5}{0}
\smallskip
\caption
{The extension-degree of supercoiling curves at
various fixed forces  for
 a model DNA.}
\label{fig5}
\end{figure}

Now we study the case when a nonzero torque $\gamma$
as well as a force $f$ are
 applied on the molecule. In this case the
DNA will be supercoiled \cite{r18,r10}.
An analysis \cite{r11}
similar with what we have done in
the above shows that, at certain fixed $f$ ($< f_c$)
 there exists a 
negative-valued critical torque $\gamma_c^{d}$,
\begin{equation}
\gamma_c^d=R (f-2.6899 \epsilon/ r_0\cos\theta_0)
\tan\theta_c,
\label{tc}
\end{equation}
if  $\gamma \leq \gamma_c^d$, the DNA will no longer be
in its native structure and it will denature 
\cite{r5,r6}. However,  there also
exists another critical
torque $\gamma_c^p\simeq\sqrt{8\cos\theta_0 f\kappa}$,
at the point where $|\gamma|=\gamma_c^p$ the 
DNA chain will begin to 
take on plectonemic configurations,
 with the total extension nearly
zero \cite{r11,r18}. Thus, for positive torques only the transition
to plectonemes is possible; while for negative torques, 
there are two possibilities: (i) 
if $\gamma_c^d<-\gamma_c^p=\gamma$,
transition to plectonemes will occur, and (ii) if
$-\gamma_c^p<\gamma_c^d=\gamma$, the DNA will denature.
The numerically calculated relations between
extension and degree of supercoiling $\sigma$ 
demonstrated in Fig. \ref{fig5} has confirmed
this theoretical picture.

From Fig. \ref{fig5} we know that, (a) at low forces 
(curves E and F), the DNA chain will change to plectonemes
at high degrees of supercoiling and the  curve
 is symmetric (achiral) 
with $\sigma\rightarrow -\sigma$;
(b) at moderate forces (curves C and D) the chirality of
DNA begins to emerge, a positively supercoiled DNA will
 shrink continuously, 
while a negatively supercoiled one will
shrink at first till
some turning point is reached,then it
will swell again, followed by   denaturation
of the double-helix;
(c) at large forces (curves A and B) the turning points
disappear, and the more negatively supercoiled is 
the chain
  the more extended  it is. All these
predictions are in excellent qualitive agreement
with the experiment of Strick {\it et al.} \cite{r5}, this
can be seen clearly by a direct comparison of
Fig. \ref{fig5} with Fig. 3 of \cite{r5}.

However, we should point out that, for supercoiled DNA,
 the quantitative
agreement is not so encouraging. Fig. \ref{fig5}
indicates negatively supercoiled DNAs will denature rather
than form plectonemes for applied forces 
$\geq 8 \;\mbox{pN}$, while  the experimental value is
just about  $1\;\mbox{pN}$ \cite{r5,r6}. 



\begin{thebibliography}{000}

\bibitem{r1}
S.\ B.\ Smith, L.\ Finzi, and  C.\ Bustamante,
 Science {\bf 258}, 1122 (1992);
C.\ Bustamante {\it et al.},
{\it ibid.} {\bf 265}, 1599 (1994).

\bibitem{r2}
J. F. Marko and 
E. D. Siggia, Macromolecules {\bf 28}, 8759 (1995).

\bibitem{r13}
N. Sait\^{o}, K. Takahashi, and Y. Yunoki, J.
Phys. Soc. Jpn {\bf 22}, 219 (1967).


\bibitem{r3}
P. Cluzel {\it et al.}, Science {\bf 271}, 792 (1996).

\bibitem{r4}
S. B. Smith, Y. Cui, and C. Bustamante, Science
 {\it 271}, 795 (1996).

\bibitem{r5}
T. R. Strick {\it et al.},
Science {\bf 271}, 1835 (1996).

\bibitem{r6}
T. R. Strick, V. Croquette, and D. Bensimon,
Proc. Natl. Acad. Sci. USA {\bf 95}, 10~579 (1998). 

\bibitem{r7}
J. K\"{a}s {\it et al.},
Europhys. Lett. {\bf 21}, 865 (1993).

\bibitem{r8}
R. Everaers, R. Bundschuh, and K. Kremer,
 Europhys. Lett. {\bf 29},
263 (1995).

\bibitem{r9}
T. B. Liverpool, R. Golestanian, and K. Kremer,
Phys. Rev. Lett. {\bf 80}, 405 (1998).

\bibitem{r10}
For a recent review of the various approaches, see
 J. F. Marko,  Phys. Rev. E {\bf 57},
2134 (1998) and references therein. 

\bibitem{r19}
H. Zhou and Z.-C. Ou-Yang, Phys. Rev. E {\bf 58},
4816 (1998); J. Chem. Phys. {\bf 110}, in the press.

\bibitem{r11}
H. Zhou and Z.-C. Ou-Yang, in preparation.

\bibitem{r12}
For actual polymers with large values of $\ell_p/R$ this
 is of little effect, and
the tangent-tangent correlation is the same as the
usual three-dimensional case, i.e., $\langle 
{\bf t}(s)\cdot {\bf t}(0)\rangle 
\sim \exp (-s/2\ell_p)$\cite{r13,r11}. 
However, in the case of $R \gg \ell_p$, because of
Eq.\ (\ref{eq1}) the tangent ${\bf t}$ 
is constrained to 
a single plane and hence 
$\langle {\bf t}(s)\cdot {\bf t}(0)\rangle 
\sim\exp(-s/4\ell_p)$ for this case \cite{r13,r11}.
 These predictions are 
different from the mean-field
results of \cite{r9}, which gives $\langle {\bf t}
(s)\cdot{\bf t}(0)\rangle \sim 
\exp(-3s/4\ell_p)$ for $R\ll \ell_p$ and 
$\langle {\bf t}
(s)\cdot{\bf t}(0)\rangle 
\sim \exp(-3s/2\ell_p)$ for $R\gg \ell_p$.
Actually  mean-field approximations may  not be
suitable for studying the tangent correlations 
[M. Otto, J. Eckert, and T. A. Vilgis,
Macromol. Theory Simul. {\bf 3}, 543 (1994);
B.-Y. Ha and D. Thirumalai, J. Chem. Phys.
{\bf 106}, 4243 (1997)].


\bibitem{r14}
The probability distribution of the  tangent
${\bf t}$ of a semiflexible has already been
 obtained \cite{r13}.

\bibitem{r15}
For example, the cell membrane  
 is a  closed
bilayered structure  of amphiphiles
 with each amphiphile molecule having 
{\it two} nonpolar hydrocarbon chains 
attached to a single polar head group.
 In room temperature 
 the bilayer is in a fluid phase,
with its
hydrocarbon chains  very flexible and much shorter
 than their full  contour length.
 However when temperature is
lowered below some threshod  a phase 
transition will occur and the membrane
 will be  in a solid phase
 with the hydrocarbons almost fully
extended.


\bibitem{r16}
W. Saenger, {\it Principles of Nucleic Acid
Structure} (Springer-Verlag, NJ, 1984).

\bibitem{r17}
 J. H. White, Am. J. Math. {\bf 91},
693 (1969); F. B. Fuller, Proc. Natl. Acad. Sci. USA
{\bf 68}, 3557 (1978).

\bibitem{r18}
J. F. Marko and E. D. Siggia, Phys. Rev. E {\bf 52},
2912 (1995).

\end{thebibliography}
\end{document}